\documentclass[lettersize, journal]{IEEEtran}
\usepackage{algorithm}
\usepackage{blindtext}
\usepackage{amsmath,amssymb,amsfonts}
\usepackage{xcolor}
\usepackage{hyperref}
\usepackage{algpseudocode}
\usepackage{caption}
\usepackage{array}
\usepackage[caption=false,font=normalsize,labelfont=sf,textfont=sf]{subfig}
\usepackage{textcomp}
\usepackage{stfloats}
\usepackage{url}
\usepackage{verbatim}
\usepackage{graphicx}
\usepackage{cite}
\usepackage{titlesec}
\usepackage{enumitem}
\usepackage{booktabs}

\titlespacing*{\section}
  {0pt}{*1.0}{*0.5}
\titlespacing*{\subsection}
  {0pt}{*1.0}{*1.0}

\title{Diffusion Model Based Resource Allocation Strategy in Ultra-Reliable Wireless Networked Control Systems}

\author{Amirhassan Babazadeh Darabi,~\IEEEmembership{Student Member,~IEEE}, Sinem Coleri,~\IEEEmembership{Fellow,~IEEE}
\thanks{A. B. Darabi and S. Coleri are with the Department of Electrical and
Electronics Engineering, Koc University, Istanbul, e-mail: {adarabi22, scoleri}@
ku.edu.tr. Sinem Coleri acknowledges the support of the Scientific and Technological
Research Council of Turkey 2247-A National Leaders Research
Grant \#121C314 and Ford Otosan.}}

\IEEEpubid{0000--0000/00\$00.00~\copyright~2021 IEEE}

\begin{document}

\maketitle

\begin{abstract}
Diffusion models are vastly used in generative AI, leveraging their capability to capture complex data distributions. However, their potential remains largely unexplored in the field of resource allocation in wireless networks. This paper introduces a novel diffusion model-based resource allocation strategy for Wireless Networked Control Systems (WNCSs) with the objective of minimizing total power consumption through the optimization of the sampling period in the control system, and blocklength and packet error probability in the finite blocklength regime of the communication system. The problem is first reduced to the optimization of blocklength only based on the derivation of the optimality conditions. Then, the optimization theory solution collects a dataset of channel gains and corresponding optimal blocklengths. Finally, the Denoising Diffusion Probabilistic Model (DDPM) uses this collected dataset to train the resource allocation algorithm that generates optimal blocklength values conditioned on the channel state information (CSI). Via extensive simulations, the proposed approach is shown to outperform previously proposed Deep Reinforcement Learning (DRL) based approaches with close to optimal performance regarding total power consumption. Moreover, an improvement of up to eighteen-fold in the reduction of critical constraint violations is observed, further underscoring the accuracy of the solution. 
\end{abstract}


\section{Introduction}
\IEEEPARstart{W}{NCS}s are control systems with control loops closed through a wireless communication network \cite{8166737}. WNCSs play an important role in supporting emerging applications in sixth-generation (6G) networks, such as remote driving \cite{10529954} and cooperative robots (cobots) \cite{10525242}. The joint optimization of the performance of the control and communication systems is the main challenge in WNCSs due to the high complexity of modeling the interactions between these two systems, the stringent ultra-reliability requirements of control systems, and the non-ideal propagation characteristics of wireless communication systems. 

Earlier research on the joint design of control and communication systems for WNCSs focused on the usage of optimization theory-based solution strategies following the derivation of the right abstractions of these two systems \cite{sadi2017joint, cao2023age}. In joint optimization frameworks that uniquely abstract the performance of control and communication systems, an optimization problem is formulated with both the control and communication decision variables and solved using iterative or heuristic methods following the derivation of the optimality conditions. However, the high complexity of the proposed model-based methods may hinder their application in low-latency WNCS scenarios. 
\IEEEpubidadjcol

In order to solve the time complexity issue of the optimization theory-based solutions and additionally operate with incomplete CSI, DRL is introduced to allocate resources optimally through trial and error in a complex scenario \cite{10197647}. In our previous work \cite{10479170}, we have proposed an optimization theory-based DRL approach, where first, the model is simplified based on the optimality conditions of the optimization problem. Then, the new simplified problem is fed to a Dueling Double Deep Q-network (D3QN) to output the optimal blocklength values. Although DRL methods use online interactions with the environment to learn, the amount of data needed to train the model is large. Moreover, DRL models may produce infeasible solutions that violate the constraints of the control and communication systems, which may have detrimental effects on critical safety applications. 

Generative AI-based approaches are considered a solution to enhance the performance of 5G/6G networks. \cite{kasgari2020experienced} introduces Generative Adversarial Networks (GANs) to be utilized along with a DRL model to enhance the model's training and adaptability to extreme conditions compared to the conventional DRL models. However, the GAN used in the proposed model is solely used to generate data to pre-train the DRL model and is not involved in the decision-making process. Additionally, GANs are known to be unstable in training time and cannot generate high-quality samples in the inference phase. To fix the instability and poor quality of the generated samples, DDPMs introduced in \cite{ho2020denoising} are proposed to be utilized in \cite{letafati2023generative} to enhance the performance of the wireless networks in constellation shaping problem. However, to this day, diffusion models are not directly applied to resource allocation problems in wireless networks. 

In this paper, we propose a novel DDPM-based resource allocation scheme for the joint design of control and communication systems in WNCSs for the first time in the literature. The DDPM is used to generate optimized blocklength values from an isotropic Gaussian distribution by using the CSI as conditional information. 

The rest of the paper is organized as follows. First, the system model is presented in Section \ref{systemodel}. Then, the proposed methodology is discussed in Section \ref{diffusion}. Performance evaluation is provided in Section \ref{performance}. Finally, the paper is concluded, and future works are discussed in Section \ref{conclusion}.

\section{System Model} \label{systemodel}
The WNCS consists of $\mathcal{N}$ sensor nodes with blocklength $m_i$, sampling period $h_i$, and packet error probability $p_i$ for $i \in \{1, 2, \dots, \mathcal{N}\}$. Sensor nodes connected to a physical plant measure and send the plant's state to a controller via a wireless channel. Based on the recent state update information, the controller decides on a new control command and sends it to the actuator to be executed. The outdated packets are not retransmitted since old state information can harm time-critical control systems. The packet error is modeled as a Bernoulli random process to simplify the problem. The Time Division Multiple Access (TDMA) method is utilized for a deterministic access delay widely preferred in various automation applications \cite{ansi2011isa}. We assume that the channel time is segmented into frames, and each frame is subdivided into time slots. The initial slot is allocated for the beacon frame, which the controller sends out periodically to disseminate synchronization and scheduling updates among the nodes within the WNCS. During the scheduling update, nodes are allocated time slots for their respective data transmissions and additional parameters, such as the optimal transmission power and blocklength. We assume that nodes within the network do not transmit simultaneously and that the network manager continuously monitors the packet error rate. 
\subsection{Optimization of Control and Communication Systems}
The joint optimization of control and communications systems for ultra-reliable communication in the finite blocklength (FBL) regime is adopted from our previous paper \cite{10479170} and is presented below.
\begin{subequations} \label{eq:pure_opt}
\begin{equation}\label{eq:1a}
\begin{aligned}
    & \underset{\substack{h_i, m_i, p_i \\ i \in [1, \mathcal{N}]}}{\text{minimize}}
    & & \sum_{i=1}^{\mathcal{N}} C_{i1} C_2 \frac{m_i}{h_i} \Big[  \exp{\Big( \frac{Q^{-1}(p_i)}{\sqrt{m_i}} + \frac{\ln{2} L_i}{m_i} \Big)} - 1 \Big]\\ & & &+ \frac{C_2 W_i^c m_i}{h_i}\\
    & \text{subject to} \\
\end{aligned}
\end{equation}
\begin{equation}\label{eq:1b}
    \begin{aligned}
    & & &\lfloor \frac{\Omega}{h_i} \rfloor \ln{p_i} - \ln{(1 - \delta)} \leq 0, \; \forall i \in [1, \mathcal{N}] \\
    \end{aligned}
\end{equation}
\begin{equation}\label{eq:1c}
    \begin{aligned}
        & & & 0 < d_i(m_i) \leq \min{(\Delta, h_i)}, \; \forall i \in [1, \mathcal{N}] \\
    \end{aligned}
\end{equation}
\begin{equation}\label{eq:1d}
    \begin{aligned}
        & & & 0 < h_i \leq \Omega, \forall i \in [1, \mathcal{N}] \\
    \end{aligned}
\end{equation}
\begin{equation}\label{eq:1e}
    \begin{aligned}
        & & & 0 < p_i \leq 1, \forall i \in [1, \mathcal{N}] \\
    \end{aligned}
\end{equation}
\begin{equation}\label{eq:1f}
    \begin{aligned}
        & & & m_i \leq M_{th}, \; \forall i \in [1, \mathcal{N}] \\
    \end{aligned}
\end{equation}
\begin{equation}\label{eq:1g}
    \begin{aligned}
        & & & C_{i1} \Big[ \exp{\Big(\frac{Q^{-1}(p_i)}{\sqrt{m_i}} + \frac{\ln{2} L_i}{m_i}}\Big) - 1 \Big] \leq W_{tx, max} \\
    \end{aligned}
\end{equation}
\begin{equation}\label{eq:1h}
    \begin{aligned}
        & & & \sum_{i = 1}^{\mathcal{N}} \frac{d_i(m_i)}{h_i} \leq \beta, \\
    \end{aligned}
\end{equation}
\end{subequations}
where $C_{i1} = \frac{\sigma^2}{|g_i|}$ and $C_2 = \frac{1}{B}$; $\sigma^2$ denotes the noise power spectral density (PSD); $g_i$ denotes the channel gain of sensor node $i$; $B$ is the bandwidth; $Q^{-1}(\cdot)$ denotes the inverse of Q function; $L_i$ is the packet length, and $W_{i}^C$ is the circuit power for node $i$. The objective function (\ref{eq:1a}) is to minimize the total power consumption in the network considering both the transmit power and the circuit power of the nodes when sending packets. Constraints (\ref{eq:1b}) and (\ref{eq:1c}) represent stochastic MATI ($\Omega$) defined as the probability of maximum allowed time interval between the reception of the state vector reports above MATI being greater than a predefined value $\delta$ and MAD ($\Delta$) defined as the maximum packet delay smaller than a maximum limit, respectively, used as an abstraction of the requirements to guarantee a certain control system performance. Constraints (\ref{eq:1d}) and (\ref{eq:1e}) give the lower and upper bounds of the sampling period and packet error probability. Equation (\ref{eq:1f}) states that plant state information is transmitted in small packets using a finite blocklength, which cannot exceed a threshold value $M_{th}$ because the transmission must finish before the maximum allowable channel uses is reached. Additionally, a maximum transmit power constraint in (\ref{eq:1g}) limits the nodes from exceeding a certain transmit power level due to the limited power source of the sensor nodes and government regulations. Moreover, the schedulability constraint (\ref{eq:1h}) ensures that transmission times are assigned to multiple sensor nodes without any two nodes transmitting simultaneously. Each node $i$ is allocated a fraction of the total schedule length, denoted as $\frac{d_i}{h_i}$. Since no two nodes can transmit simultaneously, the sum of these terms represents the total time allocated to all nodes relative to the schedule length. The problem is a non-convex Mixed-Integer programming problem, so searching for a global optimum solution is difficult \cite{boyd2004convex}. 
\raggedbottom
\subsection{Simplified Optimization Problem}
The problem in (\ref{eq:pure_opt}) is not tractable and has to be simplified in order to reach sub-optimal results. As a result, the solution is grouped into multiple blocks based on the derivation of the optimality conditions, and the decision variables are reduced to consider blocklength only. Then, the other decision variables can be obtained through optimality conditions. 

The optimality conditions are derived in \cite{10479170} as
\begin{equation} \label{eq:lemma1}
    \frac{\Omega}{h_i^*} = \frac{\ln{(1 - \delta)}}{\ln{p_i^*}} = k_i,
\end{equation}
where $k_i$ is a positive integer. Next, the optimal value of $k_i$ is derived in terms of $m_i$ as

\begin{equation}\label{eq:ki}
    \resizebox{\columnwidth}{!}{
    $
    k_i^* (m_i) = \max{\left[ 1, \: \left\lceil \frac{\ln{(1 - \delta)}}{\ln{\left[ Q\left(\sqrt{m_i} \ln{\left( \frac{W_{tx, max}}{m_i C_{i1}} + 1\right)} - \frac{\ln{(2)}L}{\sqrt{m_i}}\right)\right]}} \right\rceil \right]}.
    $
    }
\end{equation}
Then, the problem (\ref{eq:pure_opt}) is simplified to reduce the decision variables and the constraints in the problem. The model is optimized using one decision variable of blocklength $m_i$ instead of three decision variables, and the other variables are derived using the optimality conditions described in (\ref{eq:lemma1}) and (\ref{eq:ki}). The modified joint optimization problem is formulated as

\begin{subequations}\label{eq:simp_opt}
    \begin{equation}\label{eq:simp_obj}
        \begin{aligned}
            & \underset{\substack{m_i, i \in [1, \mathcal{N}]}}{\text{minimize}} & &\sum_{i=1}^{\mathcal{N}} C_{i1} C_2 \frac{m_i k_i^*(m_i)}{\Omega}\\ 
             & & & \Big[ \exp\Big( \frac{Q^{-1}\left((1 - \delta)^{\frac{1}{k_i^*(m_i)}}\right)}{\sqrt{m_i}} + \frac{\ln 2 \cdot L_i}{m_i} \Big)\\
             & & &  - 1 \Big] + \frac{C_2 W_i^c m_i k_i^*(m_i)}{\Omega}\\
            & \text{subject to}\\
        \end{aligned}
    \end{equation}
    \begin{equation}
        \begin{aligned}
            & & & m_i \leq M_{th}, \; \forall i \in [1, \mathcal{N}] \\
        \end{aligned}
    \end{equation}
    \begin{equation}
        \begin{aligned}
            & & &\sum_{i=1}^{\mathcal{N}} \frac{C_2 m_i k_i^*(m_i)}{\Omega} \leq \beta. \\
        \end{aligned}
    \end{equation}
\end{subequations}

\section{DDPM-Based Resource Allocation Algorithm} \label{diffusion}
\begin{figure}[!t]
    \centering
    \includegraphics[width= 0.9\columnwidth]{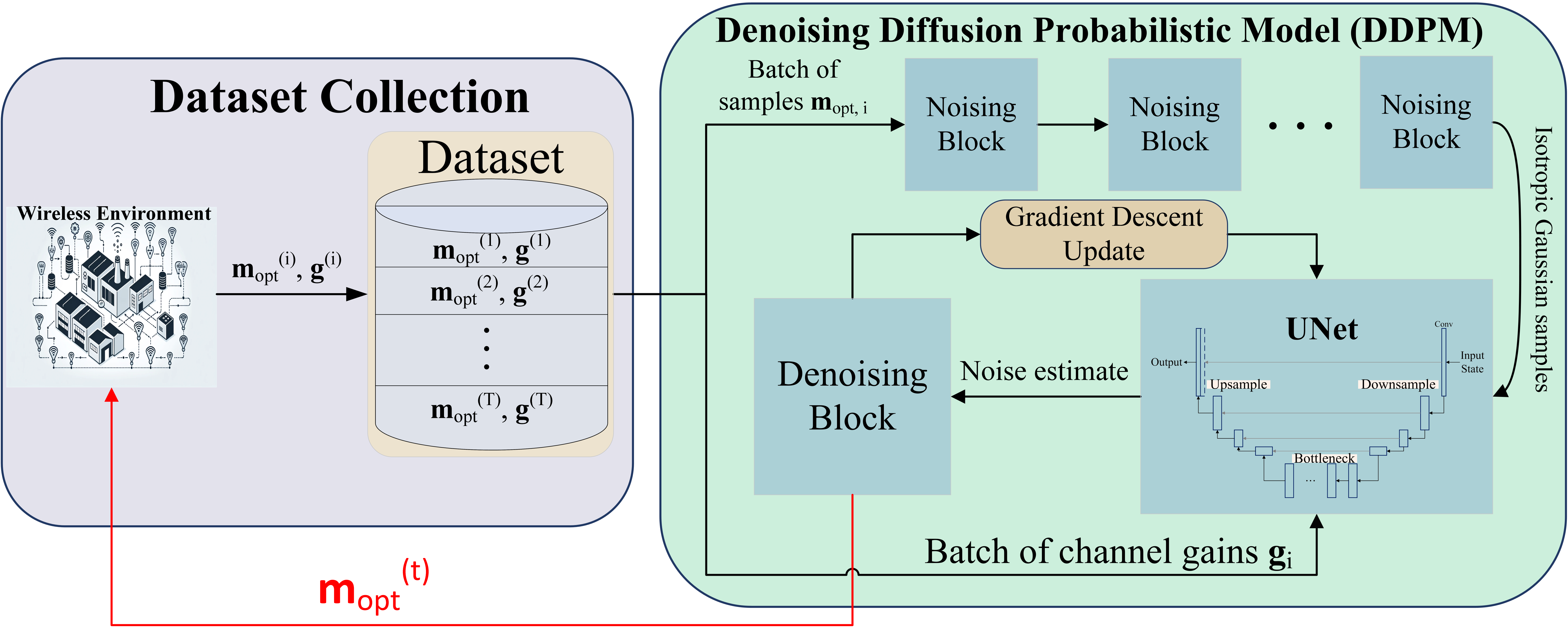}
    \captionsetup{justification=justified, singlelinecheck=false}
    \caption{The DDPM-based resource allocation algorithm.}
    \label{fig:proposed}
\end{figure}

The proposed diffusion-based resource allocation algorithm is a centrally-trained-centrally-executed model consisting of two stages: optimization theory-based data collection and DDPM-based training and data generation, as depicted in Fig. \ref{fig:proposed}. In the optimization theory-based data collection stage, various values of channel gains and the corresponding optimal blocklength values are collected. The optimal blocklength values are determined by solving the optimization problem in (\ref{eq:simp_opt}). The resulting dataset is then the input to the the diffusion model. In the diffusion model stage, the collected dataset is used to train a diffusion model and learn the optimal parameters to choose an action for blocklength adaptation for given channel gains.

The model is implemented in the controller, where control commands are sent to each sensor node after execution. DDPM consists of input states and conditional information as provided below:
\begin{itemize}[noitemsep, topsep=0pt, parsep=0pt, partopsep=0pt]
    \item \textit{Input States:} The objective of the DDPM-based method is to generate outputs drawn from a similar distribution to the input states. In the training phase, the input states are the optimal blocklength values from the dataset. Batches of data are sampled from the dataset. They are input to the model to modify and update the parameters of the neural network so the model can learn the solution space distribution and generate desired outputs after the training phase. On the other hand, in the inference phase, the input states are drawn from an isotropic Gaussian distribution with mean zero and standard deviation one, and the outputs are generated through the denoising process of the diffusion models. 
    \item \textit{Conditional Information:} The conditional information given to the network is the CSI of the links to condition the learning process on the environmental variables to ensure the model is trained to execute actions based on the current channel state. This is the most important part of the model since the environment directly affects the learning process. In the training time, in addition to the CSI, uniformly sampled time steps are given to the model to train the model to denoise the samples efficiently. 
\end{itemize} 
The proposed algorithm is summarized in Algorithm \ref{alg:proposed_diff}, which is comprised of three parts, namely, initialization and dataset collection from optimization theory-based solution, training phase, and execution phase.

\begin{algorithm}[!htb]
    \small
    \caption{Proposed Diffusion Based Resource Allocation Algorithm}\label{alg:proposed_diff}
    \begin{algorithmic}[1]
    
    \Statex \textbf{Initialization and Dataset Collection from Optimization Theory-based Solution:}
    \State Initialize network with parameters $\boldsymbol{\theta}$
    \State Collect dataset $(\mathbf{m}_{\text{opt}}, \mathbf{g})$ from optimization theory-based solution over $\mathcal{T}$ time frames
    \State Set total noising steps $T$ and variance scheduler $\beta_t$

    \Statex \textbf{Training Phase:}
    \State For each time frame:
        \State \quad Feed batch of blocklength values $\mathbf{m}_i$ as states
        \State \quad Sample time step $t \in \{1, \dots, T\}$
        \State \quad Inject noise with variance $\beta_t$
        \State \quad Pass noisy samples to model with $t$ and channel gains $\mathbf{g}_i$
        \State \quad Predict noise $\boldsymbol{\epsilon}_{\boldsymbol{\theta}}$
        \State \quad Calculate loss $\mathcal{L}_t$
        \State \quad Update network parameters
    \State Continue training until convergence

    \Statex \textbf{Execution Phase:}
    \State Pass channel gains into model during inference
    \State Predict noise and denoise to get optimal blocklength values
    \State Broadcast actions to sensor nodes

    \end{algorithmic}
\label{alg1}
\end{algorithm}

The network is initialized with parameters $\boldsymbol{\theta}$, and a dataset, which includes the optimal blocklength and channel gain values ($\mathbf{m}_{\text{opt}}, \mathbf{g}$), is collected from the environment solving the optimization problem (\ref{eq:simp_opt}). The optimization theory-based solution develops an approximation algorithm based on the analysis of the optimality conditions and the relaxation of the resulting integer optimization problem. Then, it searches for the integral solution using a greedy algorithm developed in \cite{10479170} for a duration of $\mathcal{T}$ time frames. The dataset enables the model to learn the environment and channel states so that it can generate the desired blocklength based on environmental changes. Moreover, the number of time steps $T$ and variance scheduling values ($\beta_t, t \in \{1, \dots, T\}$), which is an essential part of the DDPM algorithm, for the forward process, are determined based on a scheduling algorithm (Lines 1-3). The choice of scheduling values and time steps play a crucial role in the effective training of the proposed model and should be determined carefully based on the complexity of the model. 

After the dataset collection process, at the beginning of the training phase, since the model's objective is to generate actions for nodes to allocate resources, it samples a mini-batch of blocklength values, denoted as $\mathbf{m}_i$, as the input state and a uniformly sampled time step $t \in \{1, \dots, T\}$. The batched data are infused with Gaussian noise based on predetermined variance $\beta_t$ and fed into the model as the input states. The time step $t$ and the batch of channel gains $\mathbf{g}_i$ are given to the model as the conditional information and the time step $t$ is encoded using a positional encoding algorithm to convert discrete values into vectors to enable matrix calculus. Moreover, the channel gains batch $\mathbf{g}_i$ is normalized because small values of channel gains combined with encoded time step $t$ can be negligible, which negatively affects the learning process (Lines 4-8). The model predicts the injected noise $\boldsymbol{\epsilon}_{\boldsymbol{\theta}}$ using the input and the conditional information at that time frame (Line 9). The channel gains provided as conditional information ensure that the model becomes familiar with the actions taken based on the channel states. The loss function $\mathcal{L}_t$, which is Mean Squared Error (MSE), is used to calculate the distance between the actual noise and the predicted noise of the model, and the model is updated through backpropagation, and its performance improves until convergence (Lines 10-12). 

After the training, the execution process starts with the trained model generating high-quality outputs from an isotropic Gaussian distribution, using only the channel gains at that time frame. The predicted noise of the model based on the conditional information is utilized to denoise the isotropic samples and generate the blocklength values (Lines 13-14). After executing the actions, the controller broadcasts the blocklength values to each sensor node so that the nodes can execute the command (Line 15).

\section{Performance Evaluation} \label{performance}
In this section, the performance of the diffusion-based resource allocation technique is compared to the benchmark models, including the pure optimization theory-based model and the DRL-based models. The DRL benchmarks, Branching Deep Q-networks (BDQ), which are suitable to solve problems with multi-variable action space, are utilized to solve the optimization problem in (\ref{eq:pure_opt}) with decision variables of blocklength, sampling period, and packet error probability. The D3QN is used for the simplified optimization problem (\ref{eq:simp_opt}) with only blocklength decision variable. First, the simulation settings for the environment and the diffusion model are given in part \ref{sim}, and then, performance comparison and analysis of the proposed and benchmark methods are presented in part \ref{analysis}. 

\subsection{Simulation Setup} \label{sim}
Simulations are conducted for a network where the nodes are uniformly spread out within a circular area with a 50-meter radius, all communicating with a central controller. The path loss and shadowing effect result in large-scale fading, and it is modeled as $PL(d_i)[dB] = PL(d_0)[dB] + 10 \alpha \log{\frac{d_i}{d_0}} + Z$, where $d_i$ is the distance of node $i$ from the central controller, $PL(d_i)$ is the path loss of node $i$ at distance $d_i$ in decibels, $PL(d_0) = 35.3 \; dB$ is the path loss at the reference distance $d_0 = 1m$, path loss exponent $\alpha = 3.76$, $Z$ is a Gaussian random variable with zero mean and standard deviation equal to $4 \; dB$ corresponding to log-normal shadowing. For small-scale fading, Jake's model \cite{10479170} is used, expressed as a first-order complex Gauss-Markov process. The simulation parameters are listed in Table \ref{tab:params}.
             
\begin{table}[!t]
    \centering
    \caption{\textsc{Simulation parameters}}
    \label{tab:params}
    \footnotesize
    \renewcommand{\arraystretch}{0.9}
    \begin{tabular}{cc|cc}
        \toprule
         \textbf{Parameter} & \textbf{Value} & \textbf{Parameter} & \textbf{Value} \\
         \midrule
         B & $100$ kHz & $M_{th}$ & \scriptsize $200$ Symbols \\
         $L_i$ & $100$ bits & $\delta$ & $0.99$ \\
         $\Delta$ & $1$ ms & $\Omega$ & $100$ ms \\
         $W_{tx, max}$ & $250$ mW & $W_c$ & $5$ mW \\
         $\mathcal{N}$ & $64$ & $\sigma^2$ & \scriptsize $-174$ dBm/Hz \\
         \bottomrule
    \end{tabular}
\end{table}

The neural network used for noise prediction has a modified UNet architecture. The grid search algorithm selects the optimal hyperparameters by systematically exploring all possible combinations of their values. The simulations are performed using the PyTorch library. Our algorithm trains the reverse process of the diffusion model with one input layer, nine convolutional layers, upsample and downsample layers, and four self-attention layers. The input $\mathbf{m}$ is a vector of optimal blocklengths, and the output is the denoised version of the same vector that is refined as the network parameters are updated. As for the conditional information, noising time step $t$ is encoded using a positional encoding algorithm to convert discrete information and values into embedding vectors. The batch size is $32$ in the training phase, and the training is done for $100$ epochs. We use the AdamW optimizer \cite{loshchilov2017decoupled} with an adaptive learning rate $\alpha^{(t)}$ for stable and robust training. The initial learning rate $\alpha^{(0)}$ is $10^{-5}$. An Exponential Moving Average (EMA) method is applied to the network after a specific time in the training phase to ensure a stable weight update mechanism, where the weight factor for the EMA is set to $0.995$. For testing, we perform $50$ random initializations with $2500$ episodes and average the results. 
\subsection{Performance Comparison and Analysis} \label{analysis}
\begin{figure}[!t]
    \vspace{-2em}
    \centering
    \subfloat[]{%
        \includegraphics[width=0.45\columnwidth]{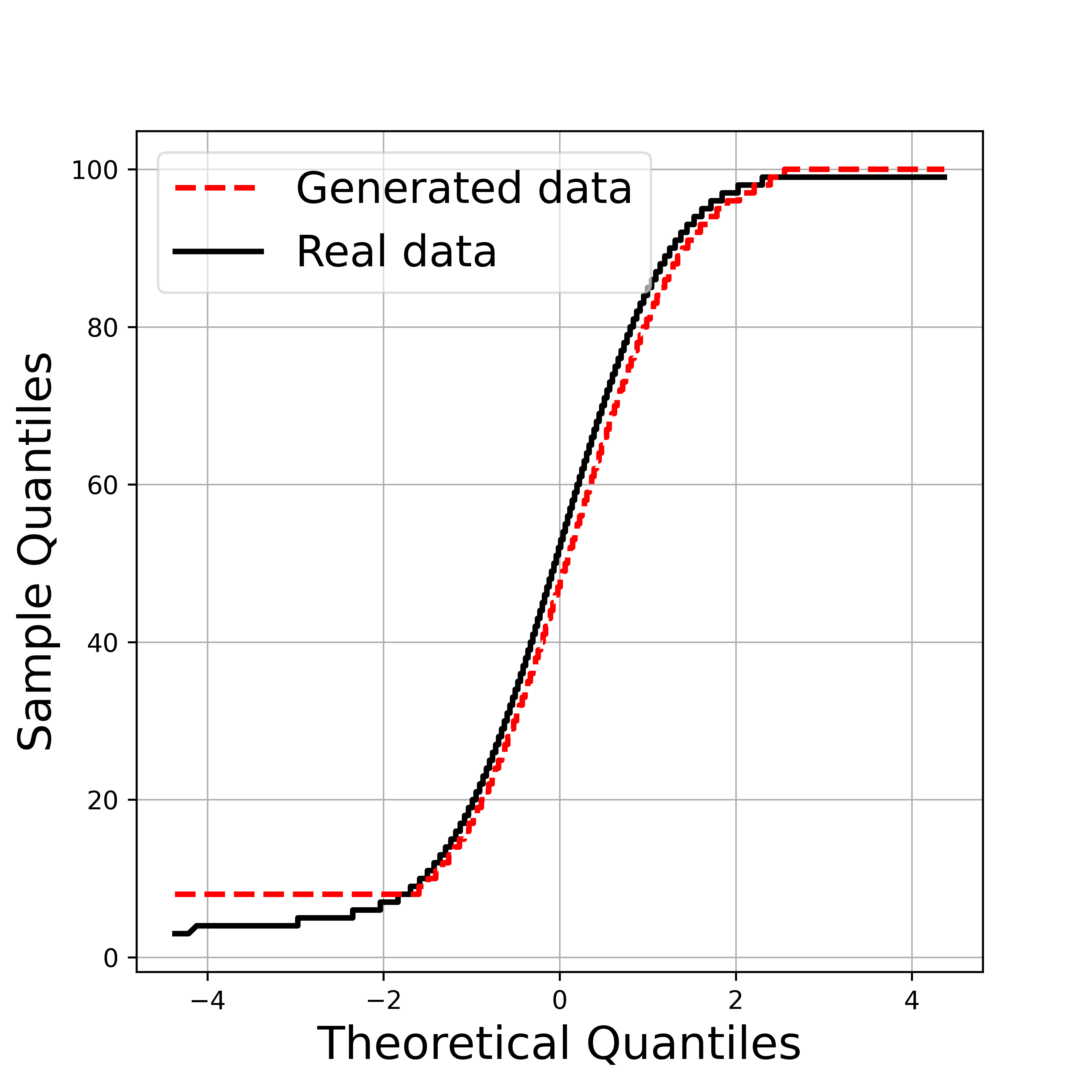} 
        \label{fig:qqplots}
    }
    \hfill
    \subfloat[]{%
        \includegraphics[width=0.45\columnwidth]{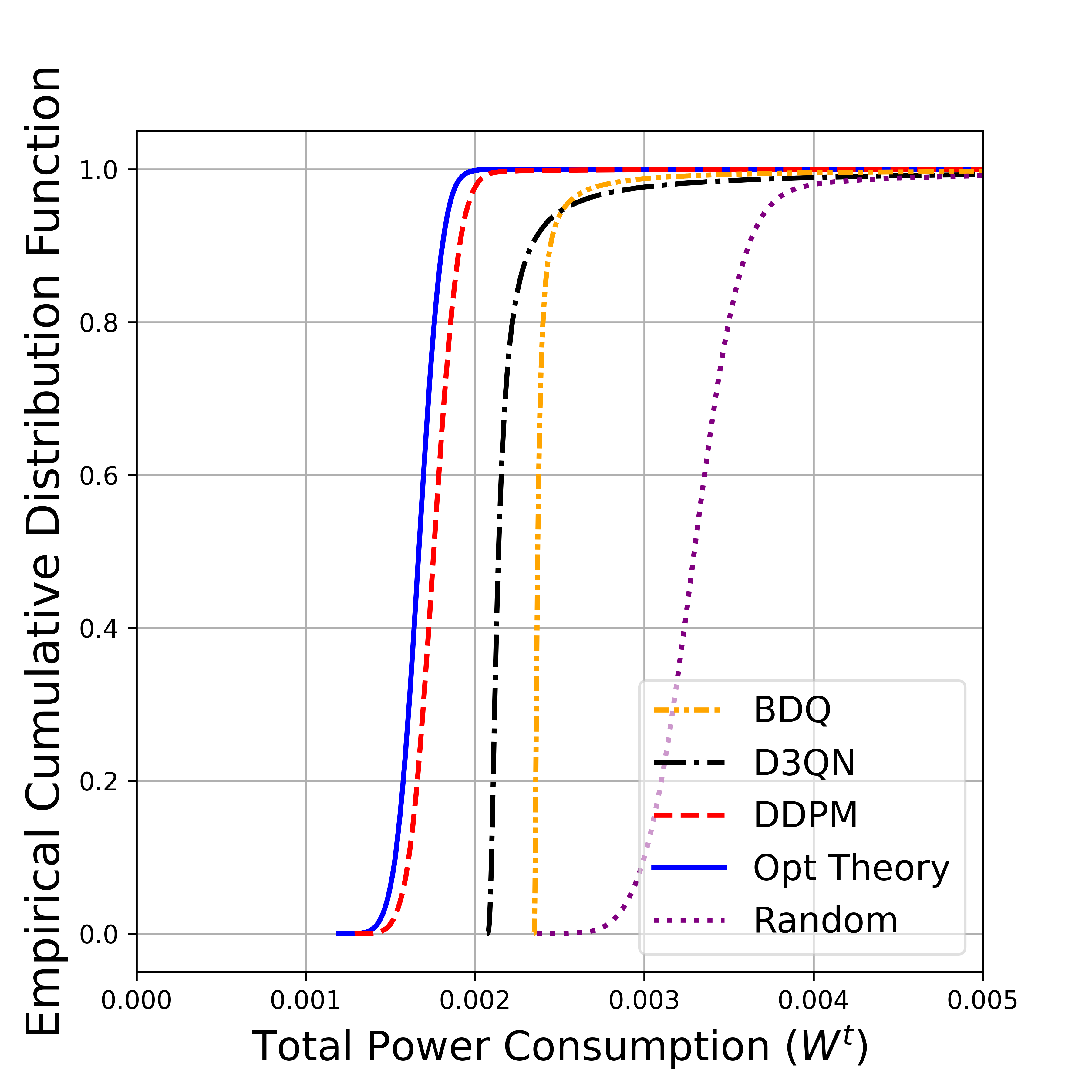} 
        \label{fig:ecdf}
    }
    \captionsetup{justification=justified, singlelinecheck=false}
    \caption{a) Q-Q plot for the true and generated samples. b) Testing results for different algorithms.}
    \label{fig:main1}
\end{figure}

We validate the accountability of the generated sample distributions compared to the original ones. The q-q plots validate the similarity between the generated and true samples. Both samples are compared to a Gaussian distribution with zero mean and unit standard deviation. Figure \ref{fig:qqplots} shows that samples exhibit similar behavior with very small discrepancy in the tail of the Gaussian distribution. This demonstrates that DDPM-based and optimization theory-based solutions are drawn from two PDFs with similar parameters.

\begin{figure}[!t]
    \vspace{-2em}
    \centering
    \subfloat[]{%
        \includegraphics[width=0.45\columnwidth]{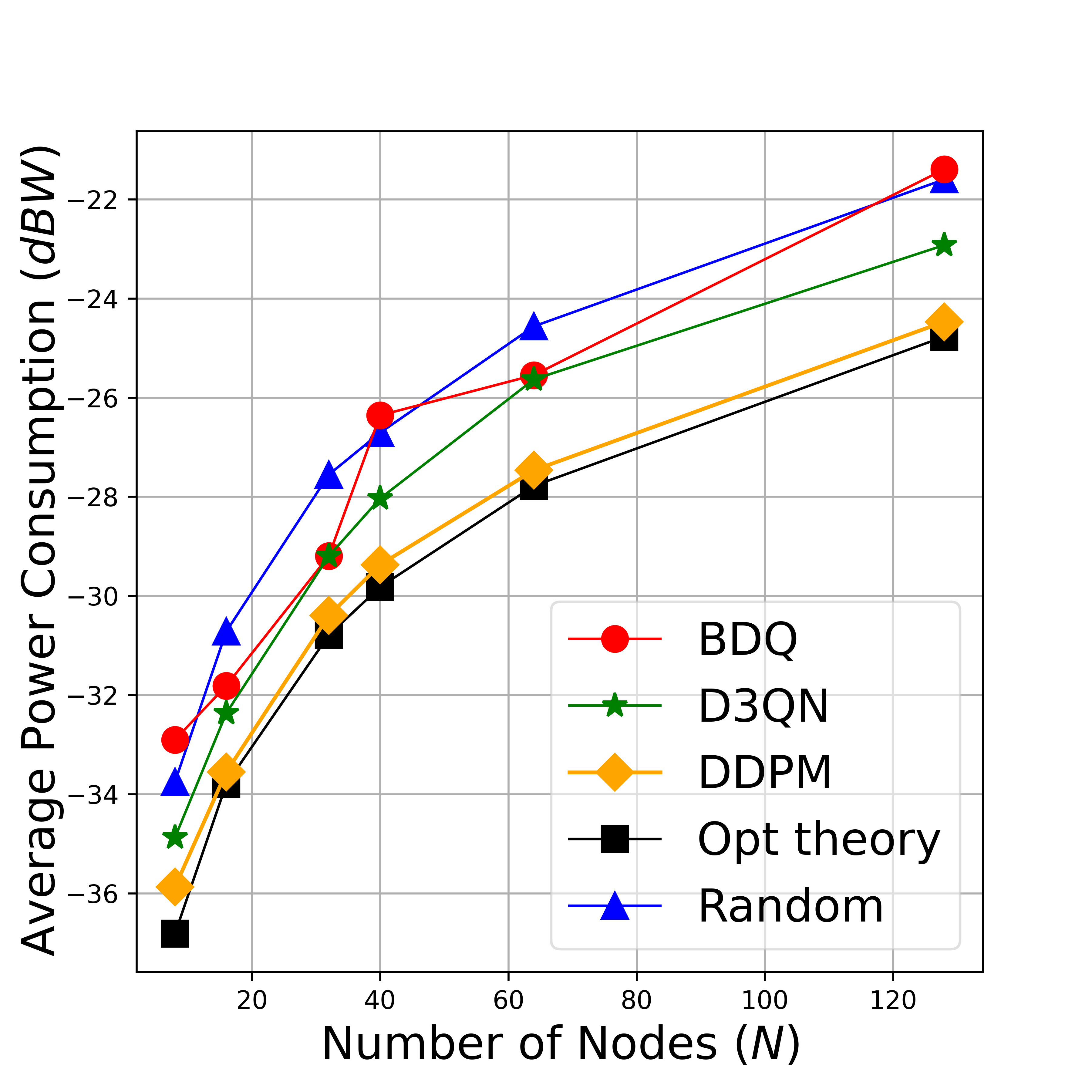} 
        \label{fig:avg_power}
    }
    \hfill
    \subfloat[]{%
        \includegraphics[width=0.45\columnwidth]{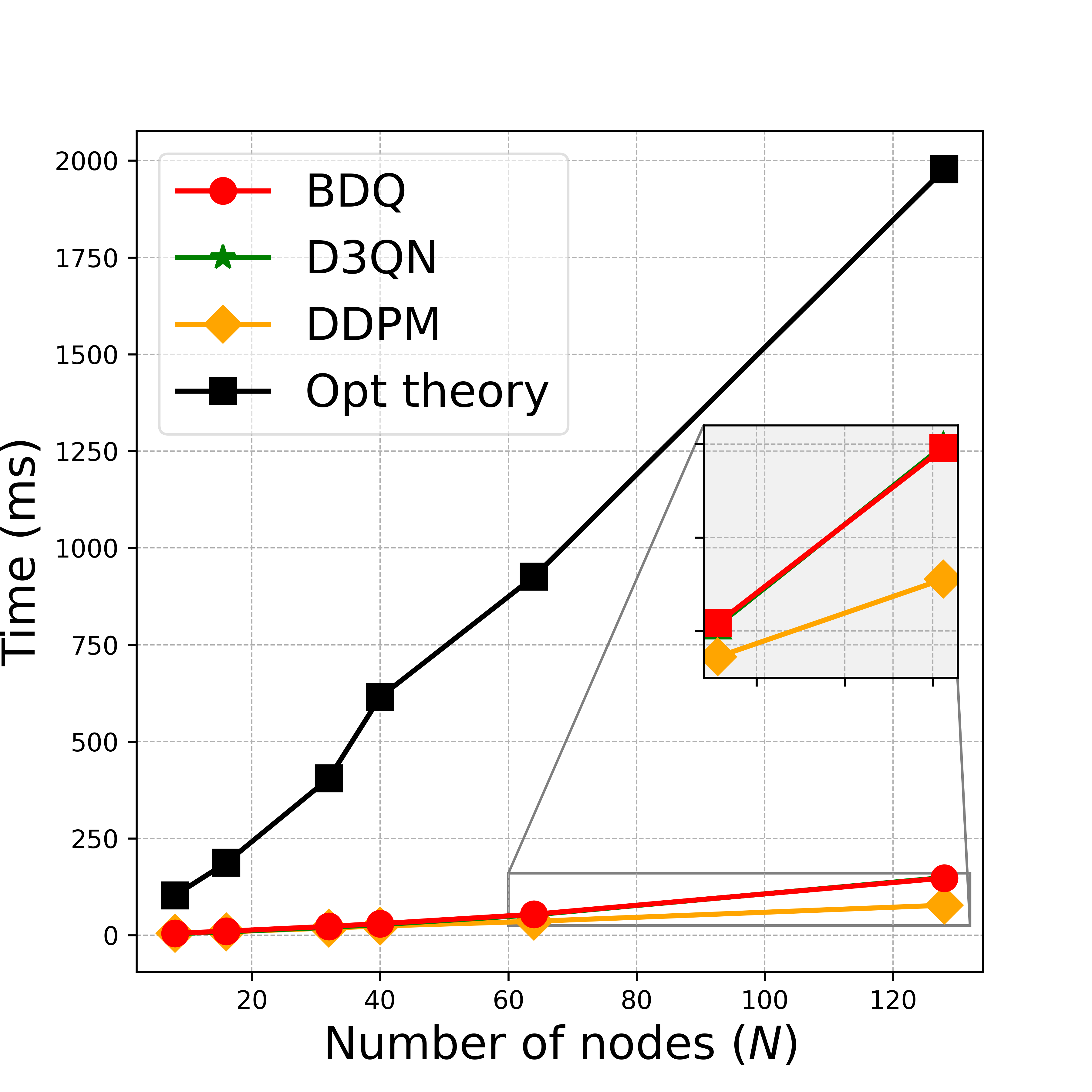} 
        \label{fig:time}
    }
    \captionsetup{justification=justified, singlelinecheck=false}
    \caption{a) Average power consumption in the testing phase. b) Average execution time for different algorithms.}
    \label{fig:main2}
    \vspace{-1em}
\end{figure}

Figures \ref{fig:avg_power} and \ref{fig:ecdf} show the average power consumption as a function of the number of nodes and the Empirical Cumulative Distribution Function (ECDF) of power consumption for 64 nodes, respectively. 
The optimization-based method demonstrates the highest performance. The DDPM-based resource allocation algorithm falls slightly short of the optimization method but is very close in performance and similarity to the optimization algorithm. As the number of nodes increases, the DDPM-based model's performance becomes similar to that of the pure optimization method. This demonstrates the scalability of diffusion models, indicating that increasing the number of nodes in the network does not result in poor performance due to the complicated structure of the network. The DRL approaches cannot surpass the DDPM-based algorithm both in average power consumption and ECDF. Finally, random selection has the poorest performance among the algorithms.

Figure \ref{fig:time} shows the execution time comparison of different algorithms as a function of the number of nodes. As expected, optimization theory-based method execution time grows exponentially as the number of nodes increases, which is not applicable in practical scenarios, especially URRLC, due to stringent conditions in the environment. On the other hand, the DDPM-based method and DRL methods exhibit linear growth as the number of nodes increases. Moreover, although the training time of the DDPM-based method is higher than the DRL-based methods due to the number of parameters in the network, the DDPM-based model shows superior performance during the testing phase. This is because the structure of the DDPM-based model allows for more effective utilization of GPU power, enabling faster performance when no further training is needed. 

\begin{figure}
    \vspace{-1em}
    \centering
    \includegraphics[width= 0.5\columnwidth]{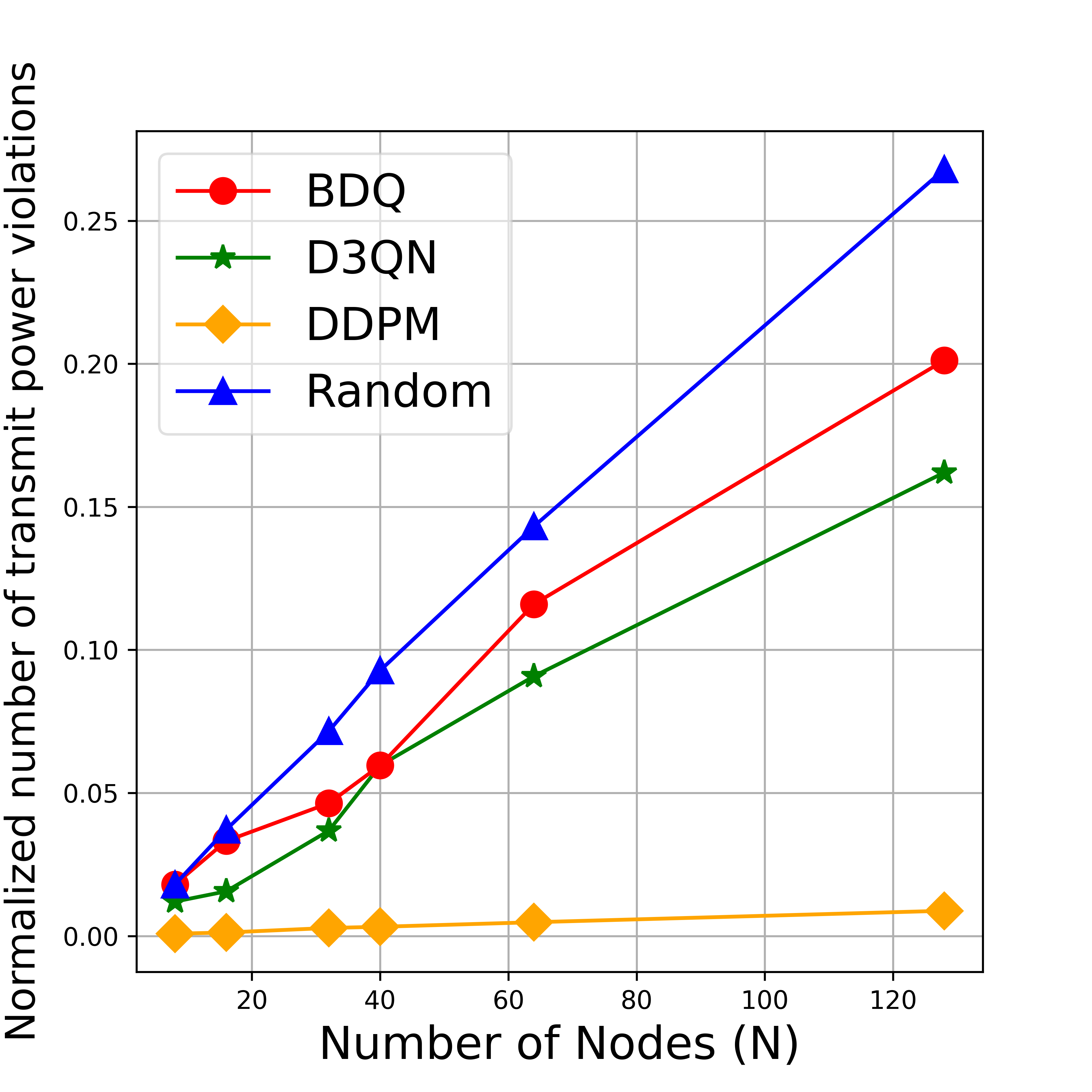}
    \captionsetup{justification=justified, singlelinecheck=false}
    \caption{Average number of violations for different algorithms as a function of the number of nodes.}
    \label{fig:violation}
    \vspace{-1em}
\end{figure}

Finally, the proposed methodology demonstrates substantial resilience in terms of violating critical constraints because it is trained to mimic the optimization theory results, but the DRL-based approaches try to avoid constraint violation by incorporating penalties in their reward function, which is not always reliable. Figure \ref{fig:violation} depicts the comparison between different algorithms for the number of times that the maximum transmit power constraint is violated as a function of the number of nodes. The result is the average number of times that algorithms have violated the power constraint in the testing phase, normalized over the total number of time steps. The proposed DDPM-based model demonstrates up to eighteen-fold improvement over cutting-edge DRL methods in terms of reliability for not violating the constraints. D3QN performs better than BDQ since it chooses actions from a smaller action space range, improving its performance and reliability. Random selection has the worst performance in terms of constraint violations.

\section{Conclusion} \label{conclusion}
In this paper, we propose a novel diffusion-based resource allocation framework for the joint optimization of communication and control systems to minimize the total power consumption of the nodes in URLLC with finite blocklength. The algorithm utilizes a DDPM model and a dataset collected from an optimization-based solution to learn the environmental variables and generate optimal blocklength values for each node. The proposed blocklength adaptation approach outperforms existing DRL-based benchmark models regarding total power consumption and performs better in avoiding actions that will cause constraint violation. In the future, we plan to investigate a DDPM-based online learning algorithm to combine the power of generative AI with DRL-based approaches where a dataset is unavailable or extremely costly to collect, like in massive MIMO communication systems. 

\bibliographystyle{ieeetr}
\bibliography{IEEE_abr,main}

\end{document}